\documentclass[
 aip,
 amsmath,amssymb,
 reprint,%
]{revtex4-2}

\usepackage{graphicx}
\usepackage{dcolumn}
\usepackage{bm}

\usepackage{natbib}
\usepackage{graphicx}
\usepackage{tikz}
\usetikzlibrary{backgrounds}
\usepackage{floatrow}
\usepackage{dcolumn}
\usepackage{bm}

\usepackage[T1]{fontenc}
\usepackage[utf8]{inputenc}
\usepackage{pgfplots}
\usepackage{grffile}
\pgfplotsset{compat=newest}
\usetikzlibrary{plotmarks}
\usetikzlibrary{arrows.meta}
\usepgfplotslibrary{patchplots}
\usepackage{amsmath}
\usepackage[utf8]{inputenc}
\usepackage[T1]{fontenc}
\usepackage{mathptmx}
\usepackage{etoolbox}
\usepackage{comment}

\begin{document}

\preprint{AIP/123-QED}

\title[Flutter Limitation of Drag Reduction by Elastic Reconfiguration]{Flutter Limitation of Drag Reduction by Elastic Reconfiguration}
\author{Maryam Boukor}
\author{Augustin Choimet}%
\author{\'Eric Laurendeau}%

\author{Fr\'ed\'erick P. Gosselin}
 \email{frederick.gosselin@polymtl.ca}
\affiliation{Department of Mechanical Engineering, Polytechnique Montr\'eal, Montr\'eal, Qc, Canada
}%

\date{8 February 2024; Published in Physics of Fluids, {https://doi.org/10.1063/5.0193649}}

\begin{abstract}

Through experiments, we idealise a plant leaf as a flexible, thin, rectangular plate clamped at the midpoint and positioned perpendicular to an airflow. Flexibility of the structure is considered as an advantage at moderate flow speed because it allows drag reduction by elastic reconfiguration, but it can also be at the origin of several flow-induced vibration phenomena at higher flow speeds. A wind tunnel campaign is conducted to identify the limitation to elastic reconfiguration that dynamic instability imposes. Here we show by increasing the flow speed that the flexibility permits a considerable drag reduction by reconfiguration, compared to the rigid case. However, beyond the stability limit, vibrations occur and limit the reconfiguration. This limit is represented by two dimensionless numbers: the mass number, and the Cauchy number. Our results reveal the existence of a critical Cauchy number below which static reconfiguration with drag reduction is possible and above which a dynamic instability with important fluctuating loads is present. The critical dimensionless velocity is dependant on the mass number. Flexibility is related to the critical reduced velocity, and allows defining an optimal flexibility for the structure that leads to a drag reduction by reconfiguration while avoiding dynamic instability. Furthermore, experiments show that our flexible structure can exhibit two vibration modes: symmetric and anti-symmetric, depending on its mass number. Because the system we consider is bluff yet aligned with the flow, it is unclear whether the vibrations are due to a flutter instability or vortex-induced vibration or a combination of both phenomena.

\end{abstract}

\maketitle

\section{Introduction}


Flexible structures in nature, from plant leaves to bird feathers, exhibit remarkable adaptive and compliant behaviours in response to aerodynamic/hydrodynamic forces \cite{vogel2020life}. Fundamental understanding of these problems can be obtained with a simplification of these structures' geometries. A flat beam or plate fixed in the middle was used to represent a plant\cite{gosselin2010drag,alben2002drag}. For small to moderate air velocities, the plate deforms statically: this mechanism is generally called reconfiguration \cite{gosselin2010drag,vogel1984drag}. Reconfiguration comes from the ability of plants to change their shape with their flexibility, which reduces the stress and allows a reduction of drag \cite{gosselin2010drag}. The more flexible the structure, the more it reconfigures, becoming more streamlined and reducing its frontal area. Several experiments have been conducted on plants to demonstrate the effectiveness of reconfiguration in reducing drag \cite{vogel2020life,de2008effects,harder2004reconfiguration,vogel1989drag}. The drag reduction has been quantified by a dimensionless number known as the Vogel exponent ($\nu$), defined as the exponent of the flow velocity in the drag formula \cite{vogel1989drag}: $F_x\propto U^{2+\nu}$, where $F_x$ is the drag force and $U$ is the flow velocity. The Vogel exponent ($\nu$) quantifies drag reduction, comparing rigid ($\nu=0$) to flexible drag ($\nu<0$) \cite{vogel1989drag}. 
Understanding fluid-structure interaction is not merely an observation of natural phenomena; it aims to extract the role of flexibility in the behaviour of these structures, obtaining fundamental principles regarding biological and physical mechanisms to inspire technological innovations and scientific advancements.

High speed flow on an elastic structure can lead to flutter—a dynamic instability resulting from competition between destabilizing aerodynamic forces and stabilizing structural stiffness \cite{leclercq2018does}. Examples of structures undergoing a flutter instability are flexible plates in axial flow \cite{eloy2007flutter,eloy2008aeroelastic}, flexible fibers \cite{rota2024forced}, flags and inverted flags \cite{kim2013flapping,tavallaeinejad2020flapping,park2019effects, padilla2024experimental}, plates in normal flow \cite{leclercq2018does}, plates under inclined flows\cite{jin2019flow}, and coupled flutter of parallel plates \cite{schouveiler2009coupled}.
Unlike a regular flag, the inverted flag exhibits flapping only within a limited range of flow velocities, and this type of response is independent of the mass number. Additionally, three-dimensional simulations have shown that the inverted flag experiences three distinct dynamic states with variations in bending rigidity and shape ratio\cite{park2019effects}.
However, in the case of a lighter inverted flag, it undergoes a fully folded dynamic mode, similar to an axial flag. \citet{sader2016large} have attributed a distinct dynamic mechanism to this specific response of the inverted flag and have related it to Vortex-Induced Vibration (VIV) and the lock-in phenomenon. The vortex-induced vibration of a structure is another mechanism of fluid/structure instability
and is presented as coupled oscillations of a solid interacting with the shed vortices in its wake \cite{blevins1977flow}. It has also been demonstrated that the dynamic response of a flexible fiber immersed in wall-bounded turbulent flow can vary based on its structural properties, exhibiting two distinct regimes: the fiber may oscillate at its natural frequency or at the turbulent frequency within the flow. More precisely, the dynamic response is linked to the ratio between these two frequencies\cite{rota2024forced}.

Here, we consider experimentally the limit to elastic reconfiguration of a plate held normal to the flow by its centre, brought by a dynamic loss of stability. A flat plate clamped at its centre and subjected to flow is used to understand the trade-off that flexibility brings to real plants in terms of drag reduction and loss of stability. We test flexible rectangular plates undergoing a two-dimensional bending deformation in a wind tunnel. Furthermore, we investigate the dynamics that occur when instability begins and we attempt to identify the mechanism responsible for the interplay between flexibility and stability that could be flutter or VIVs. A series of tests and analyses have been conducted to enhance our understanding of the physics involved in the interaction between the fluid and the flexible structure.

 \section{Methodology}\label{sec:methodology}
The experiments were conducted in the closed-loop wind tunnel (model 407B,  Engineering Laboratory Design Inc.) of the Mechanical Engineering Department at Polytechnique Montreal. The wind tunnel comprises a test section with dimensions of 61.0 $\mathrm{cm}$ (height) x 61.0 $\mathrm{cm}$ (width) x 121.9 $\mathrm{cm}$ (length). It is equipped with a 150HP motor, enabling the generation of airflow velocities ranging from 3.0 $\mathrm{m/s}$ to 91.4 $\mathrm{m/s}$. At a mean velocity of 30 $\mathrm{m/s}$, the turbulence intensity within the wind tunnel is approximately 0.21$\%$. Airflow velocity adjustments were achieved either manually by setting the wind tunnel frequency or automatically through a ramp test to assess the system's response to gradually changing flow speeds.

\begin{figure}
\centering
\normalfont
\def\svgwidth{\textwidth}
\begingroup%
  \makeatletter%
  \providecommand\color[2][]{%
    \errmessage{(Inkscape) Color is used for the text in Inkscape, but the package 'color.sty' is not loaded}%
    \renewcommand\color[2][]{}%
  }%
  \providecommand\transparent[1]{%
    \errmessage{(Inkscape) Transparency is used (non-zero) for the text in Inkscape, but the package 'transparent.sty' is not loaded}%
    \renewcommand\transparent[1]{}%
  }%
  \providecommand\rotatebox[2]{#2}%
  \newcommand*\fsize{\dimexpr\f@size pt\relax}%
  \newcommand*\lineheight[1]{\fontsize{\fsize}{#1\fsize}\selectfont}%
  \ifx\svgwidth\undefined%
    \setlength{\unitlength}{581.24176746bp}%
    \ifx\svgscale\undefined%
      \relax%
    \else%
      \setlength{\unitlength}{\unitlength * \real{\svgscale}}%
    \fi%
  \else%
    \setlength{\unitlength}{\svgwidth}%
  \fi%
  \global\let\svgwidth\undefined%
  \global\let\svgscale\undefined%
  \makeatother%
  \begin{picture}(1,0.55656937)%
    \lineheight{1}%
    \setlength\tabcolsep{0pt}%
    \put(0,0){\includegraphics[width=\unitlength,page=1]{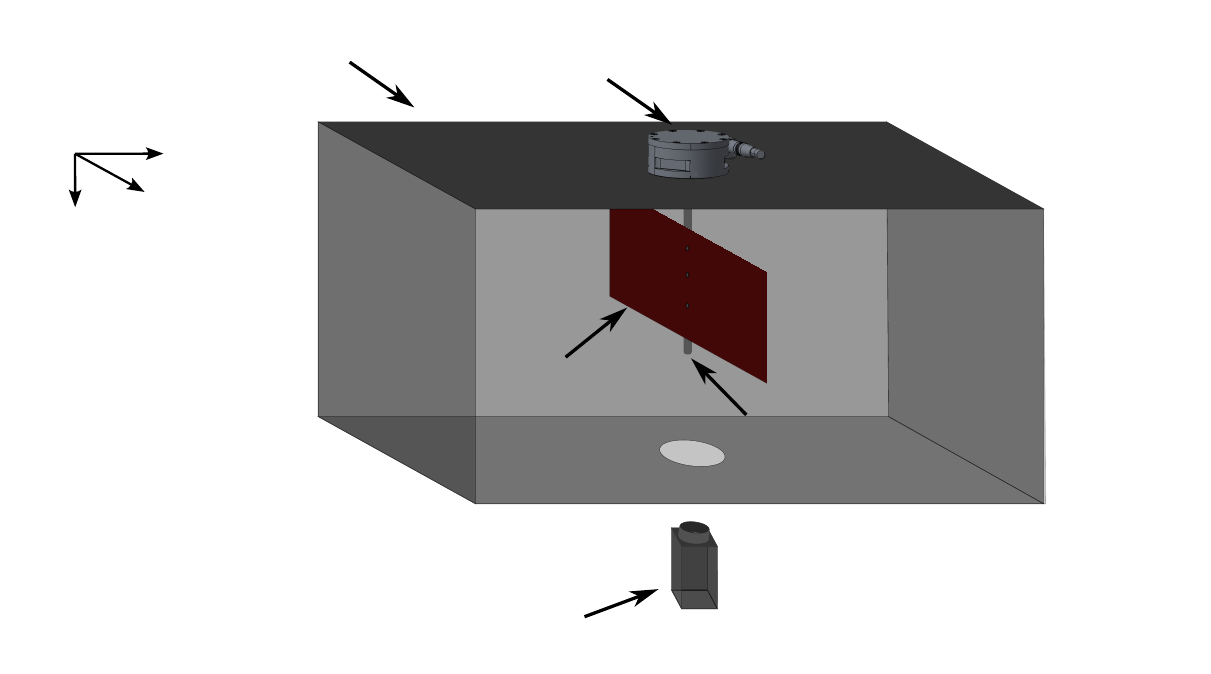}}%
    \put(0.13635235,0.41917304){\color[rgb]{0,0,0}\makebox(0,0)[lt]{\lineheight{1.25}\smash{\begin{tabular}[t]{l}x\end{tabular}}}}%
    \put(0.12119085,0.38691451){\makebox(0,0)[lt]{\lineheight{0}\smash{\begin{tabular}[t]{l}y\end{tabular}}}}%
    \put(0.05699639,0.36401098){\makebox(0,0)[lt]{\lineheight{0}\smash{\begin{tabular}[t]{l}z\end{tabular}}}}%
    \put(0.25547444,0.50000002){\makebox(0,0)[lt]{\lineheight{0}\smash{\begin{tabular}[t]{l}1\end{tabular}}}}%
    \put(0.48722631,0.51459856){\makebox(0,0)[lt]{\lineheight{0}\smash{\begin{tabular}[t]{l}2\end{tabular}}}}%
    \put(0.42518247,0.23905119){\makebox(0,0)[lt]{\lineheight{0}\smash{\begin{tabular}[t]{l}3\end{tabular}}}}%
    \put(0.62591244,0.18795622){\makebox(0,0)[lt]{\lineheight{0}\smash{\begin{tabular}[t]{l}4\end{tabular}}}}%
    \put(0.44525543,0.02919716){\makebox(0,0)[lt]{\lineheight{0}\smash{\begin{tabular}[t]{l}5\end{tabular}}}}%
  \end{picture}%
\endgroup%

\caption{A schematic representation of the experimental setup is provided. The tests were conducted within a wind tunnel (1) using a force balance (2) to measure the forces acting on a rectangular plate (3) fixed to an aluminum mast (4) at its centre. Simultaneously, a high-speed camera (5) recorded the plate's behaviour from a bottom-view perspective. }
\label{fig:setup}
\end{figure}

Figure $\ref{fig:setup}$ illustrates the experimental setup. It includes a force balance mounted flush with the test section ceiling. The tested specimens are pinned in the middle by stitching them with a steel wires to an aluminum mast in the least invasive way to minimize the use of intrusive clamps. The wire is stitched through two or three 1 mm holes, depending on the width of the plate, in the middle of the specimen to pin it to the mast. The mast is free at one end and is connected at the other end to an aluminum disk affixed to the force balance. The plate-to-mast fixation was determined to be the optimal choice for minimizing aerodynamic interference. The load cell is a Gamma type manufactured by ATI Technologies, and it has a load capacity of $30\mathrm{N}$ in the $x$ and $y$ directions and $100\mathrm{N}$ in the $z$ direction. Additionally, the system's measurements in the $F_x$ and $F_y$ directions are accurate within $\pm 0.24 \mathrm{N}$, ensuring reliability within the calibrated range of $-32 \mathrm{N}$ to $+32 \mathrm{N}$. Furthermore, the load cell exhibits exceptional sensitivity, detecting changes as small as $1/80 \mathrm{N}$ for $F_x$ and $F_y$. A LabVIEW program is used for data acquisition at a sampling frequency of $2000 \mathrm{Hz}$ with a 30-second acquisition time. 
The choice of mast dimensions is based on the plate width $w$, aiming to minimize drag resulting from the mast's frontal area while ensuring rigidity. 
The fluid load of each mast without any specimens attached is measured. The measured fluid load of a specimen is linearly corrected by subtracting the load of the exposed part of the mast during data analysis. Additionally, experimental characterization of the masts was performed to allow analyzing the specimen dynamics with more confidence, see Table \ref{tab:mast natural frequency} Appendix 1. Another experiment was conducted to test the rigid case using acrylic plates with the same dimensions as the flexible one tested, and under the same conditions as the flexible case.

A CO$_2$ laser cutter (Speedy 400, Trotec)  cut out the flat plate specimens  to the desired dimensions. A scale measured the mass of the feedstock polymer sheets. We obtain the deflection of the strips of sheets material bending under its own weight to obtain its flexural rigidity $D$. Each specimen is defined by a mass number representing the ratio of fluid mass to structure mass: 
\begin{equation}
M^*=\rho_f L/m_s,    
\end{equation}
where $\rho_f$ is the air density, $L$ is the length of the specimen, and $m_s$ is the surface density. These specimens properties are directly influenced by the specific type of polyester material chosen for fabrication. Detailed informations regarding the characteristics of each tested specimen can be found in Table \ref{tab:specimen characteristic} Appendix 1. The distance between the specimen and the tunnel ceiling is adjusted based on the size of the specimen and the mast. For smaller specimens, it is nearly the width of the specimen, for medium-sized specimens, it ranges from 1/2 to 1 width, and for larger specimens, it varies from 1/5 to 4/5 of the specimen's width.

The wind tunnel testing procedure involves employing three methods to acquire all the necessary information: extracting forces using a load cell, obtaining frequencies and identifying vibration modes through image and video acquisition using a camera setup (Blackfly 20S4C-C, Flir), and conducting visual observations.
Each specimen undergoes two tests. The first test involves incrementally increasing the flow velocity by 1.5 m/s to determine the range of critical velocities corresponding to the plate's initial vibration onset and transitions between vibration modes. The second test is conducted with a reduced velocity increment of 0.3 m/s, specifically within the critical velocity intervals.

Regarding force acquisition, the load cell is employed to detect the deformation experienced in the material's strain gauge when subjected to an external force. 
To determine the critical velocity associated with the onset of vibrations and the transition from static to dynamic mode, the coefficient of variation $C_{v_x}=\sigma/\bar{F}_x$ is computed as the ratio between the standard deviation $\sigma$ and the mean drag $\bar{F}_x$ for each tested flow velocity. This statistical approach serves to identify substantial variations within the data that could signal a critical transition.

A set of images and videos was captured during wind tunnel testing using image processing camera, equipped with both automatic and manual controls. Image capture was conducted at various flow speeds, particularly focusing on critical speeds. The camera was positioned beneath the wind tunnel, allowing for a bottom view of the setup through a circular hole in the floor, precisely aligned with the mounting of the mast and specimen. This hole was sealed with a plexiglass plug, securely fastened to ensure necessary airtightness and prevent any unwanted interference with the airflow. The Flir camera was chosen for its maximum frames per second capability (FPS=175 $\mathrm{Hz}$), and its Spinnaker software was used for camera configuration. Settings such as frame rate, exposure time, acquisition duration, and resolution were directly adjusted within the software. 
To optimize the brightness of the captured images and enhance visibility in diverse lighting conditions, LED panels were installed on both sides of the test section.

\section{Results}
\subsection{Elastic reconfiguration and dynamical occurence}

Figure \ref{fig:pictures} presents a sequence of photos of the progressive elastic reconfiguration of a flexible plate under uniform air flow (Multimedia available online). For increasing flow speed, the plate bends more and more, aligning more closely with the airflow, until a critical speed is reached. At this stage, an interesting transition occurs, marked by symmetrical and then anti-symmetrical vibrations of the plate. This preliminary observation suggests a correlation between the structure's dynamics and the increase in flow velocity, even before analyzing the aerodynamic forces obtained from the tests. To illustrate this, we analyze Figure \ref{fig:Fx vs Cvx Vs U Mylar}a, which shows the variation of drag with airspeed for two plates with the same dimensions but different materials: one being a rigid acrylic case, and the other a flexible case My140. The  markers in Figure \ref{fig:Fx vs Cvx Vs U Mylar}a represent the average drag of the flexible specimen, whereas the error bars indicate the $97^{\mathrm{th}}$ and $3^{\mathrm{rd}}$ percentile. The $97^{\mathrm{th}}$ percentile informs us of the maximal loads facing the structure, without being too sensitive to outlier measurements. See Appendix 1.

\begin{figure*}
\centering
\normalfont
\def\svgwidth{0.8\textwidth}
\input{figure2.tex}
\caption{Typical deformation of a rectangular plate subjected to flow. Bottom view photographs of the deformation of specimen Yu140. This specimen deforms statically when subjected to flow velocities of: a) 0; b) 1.7; c) 5.7; and d) 9.0 $\mathrm{m/s}$; exhibits symmetrical flutter under flow velocities of: e) 10.7; f) 10.7; g) 11.2; h) 12.7; i) 12.7; and j) 12.7 $\mathrm{m/s}$; and anti-symmetrical flutter at flow speeds of k) 14.2; and l) 14.2 $\mathrm{m/s}$. Scale bar is $4~\mathrm{cm}$ long. (Multimedia available online)}
\label{fig:pictures}
\end{figure*}

Figure \ref{fig:Fx vs Cvx Vs U Mylar}a clearly shows that the drag force increases with velocity for both the rigid and flexible cases, but the curves are different between the two. Figure \ref{fig:Fx vs Cvx Vs U Mylar}a also provides an inset focusing into a specific flow range (between 0 $\mathrm{m/s}$ and 16 $\mathrm{m/s}$) taken from the first figure. This zoom allows for a closer examination of the transition between the static and dynamic behaviour of the flexible plate. In the rigid case, drag increases following a law $U^2$ with velocity. Whereas in the flexible case, it increases in a less pronounced way, resulting in a drag reduction compared to the rigid case. This difference is crucial for understanding the impact of flexibility on drag. As the velocity surpasses 16 $\mathrm{m/s}$, we observe a sudden increase in the amplitude of the drag force, along with the maximum flexible drag surpassing the rigid drag. This means that the advantage of flexibility, which initially reduced drag, is now reversed due to the onset of flutter instability.

\begin{figure}
\centering
\normalfont
\def\svgwidth{\textwidth}
\input{figure3.tex}
\caption{Comparison of the drag between flexible and rigid plates, along with the identification of the critical velocities. a) The average drag measurements of a flexible plate My140 ($\textcolor[rgb]{0.6,0.22,0.22}{\square}$) are presented, along with the maximum and minimum values ($\textcolor[rgb]{0.6,0.22,0.22}{\rule{0.3cm}{0.5mm}}$), representing the $97^{\mathrm{th}}$ and $3^{\mathrm{rd}}$ percentiles, respectively. These measurements are plotted as a function of flow velocity and compared with drag measurements of a rigid plate ($\textcolor[rgb]{0,0,0}{\rule{0.5cm}{0.5mm}}$) that shares the same dimensions. A close-up of the graph is provided in the inset; b) Coefficient of variation applied to the drag force data. (I) First vibration mode: symmetrical. (II) Second vibration mode: anti-symmetrical.}
\label{fig:Fx vs Cvx Vs U Mylar}
\end{figure}

Moving to Figure \ref{fig:Fx vs Cvx Vs U Mylar}b, we delve into an intriguing facet of our analysis. This section of the figure illustrates the coefficient of variation applied to the drag force data as a function of velocity. This will enable us to pinpoint the instability velocity, corresponding to the transition from the static regime with reconfiguration to the dynamic regime, as well as the shift from the first flutter mode to the second. For instance, the critical velocity for the depicted case ($M^*=0.73$) is 16.2$\pm$0.15 $\mathrm{m/s}$. It is imperative to emphasize that the critical velocity identified through this statistical approach agrees with visual observations during the experiments. This approach also allows us to detect the transitions between vibration modes. In this test, we encountered two flutter modes. The first is a symmetric mode (Figure \ref{fig:pictures} e-j) that manifests initially with the onset of flutter, and the second is observed at 21.8 $\mathrm{m/s}$, displaying an anti-symmetric dynamics (Figure \ref{fig:pictures} k-l).

We then proceed to explore another specimen, the results of which are presented in Figures $\ref{fig:Fx vs Cvx Vs U Blue plastic}$a and $\ref{fig:Fx vs Cvx Vs U Blue plastic}$b, distinguished from the first by its higher mass number, $M^* = 2.87$. Comparing the results between My140 and Bl400, several similarities emerge, including the behaviour of flexible drag with air speed in comparison to rigid drag, leading us to the same conclusion as in the first case. In this second scenario, the critical velocity is approximately $3.18\pm0.15$ $\mathrm{m/s}$. We also note that the onset of flutter is accompanied by the anti-symmetric mode, as visually observed. No symmetric mode of instability is observed for this specimen. The comparison between these two cases allows us to highlight the influence of mass number on the appearance of flutter and the dynamic response of the flexible structure. We can note that the observed instability is marked by an amplification in amplitude, transitions in dynamic modes, and responsiveness to the mass number.

\begin{figure}
\centering
\normalfont
\def\svgwidth{\textwidth}
\input{figure4.tex}
\caption{Comparison of the drag between flexible and rigid plates, along with the identification of the critical velocity. a) The average drag measurements of a flexible plate Bl400 ($\textcolor[rgb]{0.6,0.22,0.22}{\square}$) are presented, along with the maximum and minimum values ($\textcolor[rgb]{0.6,0.22,0.22}{\rule{0.3cm}{0.5mm}}$), representing the $97^{\mathrm{th}}$ and $3^{\mathrm{rd}}$ percentiles, respectively. These measurements are plotted as a function of flow velocity and compared with drag measurements of a rigid plate ($\textcolor[rgb]{0,0,0}{\rule{0.5cm}{0.5mm}}$) that shares the same dimensions. b) Coefficient of variation applied to the drag force data. First vibration mode (symmetrical) is not observed. (II) Second vibration mode: anti-symmetrical.}
\label{fig:Fx vs Cvx Vs U Blue plastic}
\end{figure}

\subsection{Frequency analysis}

After characterizing the vibrational behaviours of the first two specimens, My140 and Bl400, we now turn our attention to a closer examination of the appearance of each vibration mode through spectral analysis of experimental data. Figure \ref{fig:Fft} presents the Fast Fourier Transform (FFT) of the drag force and lift force data for the specimen Yu140.

\begin{figure}
\centering
\normalfont
\def\svgwidth{\textwidth}
\input{figure5.tex}
\caption{Investigation of frequency distribution in drag ($\textcolor[rgb]{0.6,0.22,0.22}{\rule{0.5cm}{0.5mm}}$) and lift ($\textcolor[rgb]{0,0,0}{\rule{0.5cm}{0.5mm}}$) forces obtained from the wind tunnel experiments at different flow velocities for the Yu140 specimen exhibiting: a) The first vibration mode (symmetrical) \textit{U}=13.3 $\mathrm{m/s}$; b) The second vibration mode (anti-symmetrical) \textit{U}=14.2 $\mathrm{m/s}$.}
\label{fig:Fft}
\end{figure}

It is worth noting that the flutter in this case initiates at $10.3 \pm 0.15$ $\mathrm{m/s}$, accompanied at first by a symmetrical vibration mode. Our focus here is on the frequency response pertaining to the first and second observable vibration modes. Indeed, in Figure \ref{fig:Fft}a, obtained at an airflow velocity of $U=13.3$ $\mathrm{m/s}$, we observe a pronounced peak around 42 $\mathrm{Hz}$ in the drag force spectrum. Upon further analysis, a significant transition to the anti-symmetrical mode is identified at an airflow velocity of $U=13.6$ $\mathrm{m/s}$. Figure \ref{fig:Fft}b showcases this second vibration mode at 14.2 $\mathrm{m/s}$, where the FFT of the lift force exhibits a larger peak compared to that of the drag force at a frequency around 35 $\mathrm{Hz}$. Vibration frequency values were confirmed through manual tracking of the $x-$ and $y-$displacements at the tip of the tested plate, based on video images of wind tunnel experiments.

To ensure the accuracy of our frequency analyses, we estimated the vortex shedding frequency from the mast holding the specimen as $f_{vsm} = U St /d$, where the Strouhal number of a cylinder is assumed roughly constant $St=0.2$, $U$ is the flow velocity, and $d$ is the mast diameter. The estimated values of $f_{vsm}$ are found to be considerably higher ($>100$ $\mathrm{Hz}$ at the lowest flow velocities) than the frequencies identified through the signal of drag and lift forces in Figure \ref{fig:Fft}. This verification supports the conclusion that the observed peaks in the frequency spectra are indeed related to the flow-induced vibration of the specimen.

\begin{figure}
\centering
\normalfont
\def\svgwidth{\textwidth}
\begingroup%
  \makeatletter%
  \providecommand\color[2][]{%
    \errmessage{(Inkscape) Color is used for the text in Inkscape, but the package 'color.sty' is not loaded}%
    \renewcommand\color[2][]{}%
  }%
  \providecommand\transparent[1]{%
    \errmessage{(Inkscape) Transparency is used (non-zero) for the text in Inkscape, but the package 'transparent.sty' is not loaded}%
    \renewcommand\transparent[1]{}%
  }%
  \providecommand\rotatebox[2]{#2}%
  \newcommand*\fsize{\dimexpr\f@size pt\relax}%
  \newcommand*\lineheight[1]{\fontsize{\fsize}{#1\fsize}\selectfont}%
  \ifx\svgwidth\undefined%
    \setlength{\unitlength}{267.08756334bp}%
    \ifx\svgscale\undefined%
      \relax%
    \else%
      \setlength{\unitlength}{\unitlength * \real{\svgscale}}%
    \fi%
  \else%
    \setlength{\unitlength}{\svgwidth}%
  \fi%
  \global\let\svgwidth\undefined%
  \global\let\svgscale\undefined%
  \makeatother%
  \begin{picture}(1,0.6741177)%
    \lineheight{1}%
    \setlength\tabcolsep{0pt}%
    \put(0,0){\includegraphics[width=\unitlength,page=1]{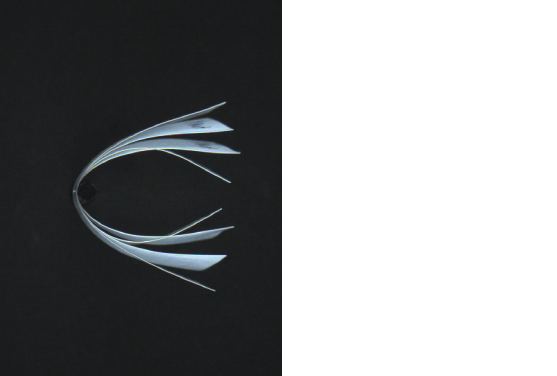}}%
    \put(0.00967467,0.61){\makebox(0,0)[lt]{\lineheight{1.25}\smash{\begin{tabular}[t]{l}\textcolor{white}{ 4 cm}\end{tabular}}}}%
    \put(0,0){\includegraphics[width=\unitlength,page=2]{figure6pdf.pdf}}%
    \put(0.34185801,0.64){\color[rgb]{0,0,0}\makebox(0,0)[lt]{\lineheight{1.25}\smash{\begin{tabular}[t]{l}\textcolor{white}{ a)}\end{tabular}}}}%
    \put(0,0){\includegraphics[width=\unitlength,page=3]{figure6pdf.pdf}}%
    \put(0.87124275,0.64){\color[rgb]{0,0,0}\makebox(0,0)[lt]{\lineheight{1.25}\smash{\begin{tabular}[t]{l}\textcolor{white}{ b)}\end{tabular}}}}%
  \end{picture}%
\endgroup%

\caption{Composite Lissajous images combining the maximum intensity of snapshots taken over the cycle of oscillation super imposed with the Lissajous traced by the superior free-end of plate YU140 at: a) $U=12.7 \mathrm{m/s}$, symmetrical mode; b) $U=15.7 \mathrm{m/s}$, anti-symmetrical mode.}
\label{fig:lissajous}
\end{figure}

Figure \ref{fig:Fft} also indicates a decrease in the fundamental frequency of the plate, dropping from 42 $\mathrm{Hz}$ to 35 $\mathrm{Hz}$, as it transitions from a symmetrical to an anti-symmetrical flutter. This can possibly be attributed to the boundary conditions of the plate. Since the plate is pinned to a mast using steel wires, the effective rigidity of the plate will change as it transitions from a symmetrical to an anti-symmetrical mode. In the symmetrical mode, as pictured in Figure \ref{fig:lissajous}a, no rotation occurs at the centre as in a clamped plate. In the anti-symmetrical case in Figure \ref{fig:lissajous}b, rotation occurs at the mast, which is expected of a pinned boundary condition. This rotation at the centre is expected to lead to a  reduction in the natural frequency of the plate.

In the previous paragraph, we defined the frequency peaks between the drag and lift signals to differentiate between symmetrical and anti-symmetrical vibration modes based on the dominant force. Examining the corresponding Lissajous Figures \ref{fig:lissajous}a and \ref{fig:lissajous}b, they reflect the characteristics of symmetrical and anti-symmetrical behaviour. Alongside visual observations and frequency analysis, this further reinforces the validity of our flutter modes classification.

Turning our focus away from the spectral analysis discussed earlier, we consider spectrogram analysis. The data used here was obtained from a ramp test on the same specimen presented in Figures \ref{fig:Fft} and \ref{fig:lissajous}. Figure \ref{fig:spectrogram yupo-14-7} illustrates a spectrogram where the abscissa corresponds to the flow velocity, while the ordinate represents the vibration frequency obtained from the experimental data of drag and lift force. The velocity is slowly ramped from 6.8 to 20.3 $\mathrm{m/s}$ over 9 minutes, while the loads from the balance are acquired continuously. The natural frequency is calculated analytically to compare it with the frequencies found in the spectrogramm, by using the following formula:
\begin{equation}
f_{nat} = \frac{a_1}{(L/2)^2} \sqrt{\frac{D w}{M+m}},
\end{equation}
where $a_1 = 3.49$ for the first mode of a clamped-free plate\cite{ballal1966vibrations}, $w$ is the width of the plate , $m$ is the linear mass, and $M$ is the linear added mass, i.e., $M=\frac{1}{4}\rho_f \pi w^2$. For Yu140 case, the natural frequency is equal to 42.7 $\mathrm{Hz}$. Firstly, we observe a band with a peak that starts around $U \approx 10.5 \mathrm{m/s} $, that clearly defines the dominance of a frequency $f_1$ equal to the system's natural frequency and corresponds to the occurrence of flutter instability. Secondly, we notice a sharp change at a flow velocity of $U \approx 13.6$ $\mathrm{m/s}$, which corresponds to the transition from the symmetric vibration mode to the anti-symmetric mode. Thirdly, we find other lines corresponding to the $2^{nd}$ and $3^{rd}$ harmonics ($f_2,f_3$), implying the existence of a nonlinear response from a fluid-structure interaction phenomenon. Linking this observation to the FFT analysis in Figure \ref{fig:Fft}a, the identified frequency $f_1$ at an airflow velocity between 10.5 $\mathrm{m/s}$ and 13.6 $\mathrm{m/s}$ in the spectrogramm aligns with the higher peak at $42 \mathrm{Hz}$ in the drag force spectrum. 
At a speed greater than 13.6 $\mathrm{m/s}$, the prominent peak in frequency corresponds to that associated with the lift signal (Figure \ref{fig:spectrogram yupo-14-7}b). It is lower than $f_1$, as we had to conclude from Figure \ref{fig:Fft}. Following this new fundamental frequency $f'_1$, it is also accompanied by the second and third harmonics ($f'_2,f'_3$), which are more clearly visible in the drag spectrogram (Figure \ref{fig:spectrogram yupo-14-7}a). Notably, the frequencies associated with the harmonics match with the findings in the FFT analysis Figure \ref{fig:Fft}.

\begin{figure}
\centering
\normalfont
\def\svgwidth{\textwidth}
\input{figure7.tex}
\caption {Representation of the dynamic behaviour of the plate Yu140 during wind tunnel tests. The spectrograms display the characteristics of frequencies (on the y-axis) of drag force (a) and lift force (b) as functions of flow velocity (on the x-axis). The color intensity represents the magnitude of the frequency.  Dashed lines represent fundamental ($f_1$), second harmonic ($f_2$), third harmonic ($f_3$), varied fundamental ($f'_1$), and harmonics of the variation ($f'_2$), ($f'_3$).}
\label{fig:spectrogram yupo-14-7}
\end{figure}

\subsection{Optimal flexibility and stability limit}
The understanding of the effect of the flexibility on the reconfiguration or the impact of the structure’s characteristics on the stability requires an investigation of the parameters involved following the dimensional analysis previously used by \citet{gosselin2010drag} to quantify the effect of flexibility on the drag of a rectangular plate. Here, we will be using the two following dimensionless numbers:
\begin{equation}
    C_y=C_D \frac{\rho_f U^2 L^3}{16 D} , \quad R= \frac{F_x}{\frac{1}{2}C_D\rho_f U_{\infty}^2 A},
\end{equation}
where $C_D$ is the drag coefficient of the equivalent rigid plate and $A=Lw$ is the surface area. The Cauchy number $C_y$ reveals the magnitude of the deformations due to dynamic pressure, and it is written as the ratio of the aerodynamic bending load to the restitutive force of the structure. On the other hand, the reconfiguration number consists in measuring the ratio between the drag of the flexible structure and the rigid drag. Rigid drag is determined through wind tunnel tests on rigid models by extracting the drag coefficient for various dimensions.

\begin{figure}
\centering
\normalfont
\def\svgwidth{\textwidth}
\begingroup%
  \makeatletter%
  \providecommand\color[2][]{%
    \errmessage{(Inkscape) Color is used for the text in Inkscape, but the package 'color.sty' is not loaded}%
    \renewcommand\color[2][]{}%
  }%
  \providecommand\transparent[1]{%
    \errmessage{(Inkscape) Transparency is used (non-zero) for the text in Inkscape, but the package 'transparent.sty' is not loaded}%
    \renewcommand\transparent[1]{}%
  }%
  \providecommand\rotatebox[2]{#2}%
  \newcommand*\fsize{\dimexpr\f@size pt\relax}%
  \newcommand*\lineheight[1]{\fontsize{\fsize}{#1\fsize}\selectfont}%
  \ifx\svgwidth\undefined%
    \setlength{\unitlength}{636.60092331bp}%
    \ifx\svgscale\undefined%
      \relax%
    \else%
      \setlength{\unitlength}{\unitlength * \real{\svgscale}}%
    \fi%
  \else%
    \setlength{\unitlength}{\svgwidth}%
  \fi%
  \global\let\svgwidth\undefined%
  \global\let\svgscale\undefined%
  \makeatother%
  \begin{picture}(1,1.06349211)%
    \lineheight{1}%
    \setlength\tabcolsep{0pt}%
    \put(0,0){\includegraphics[width=\unitlength,page=1]{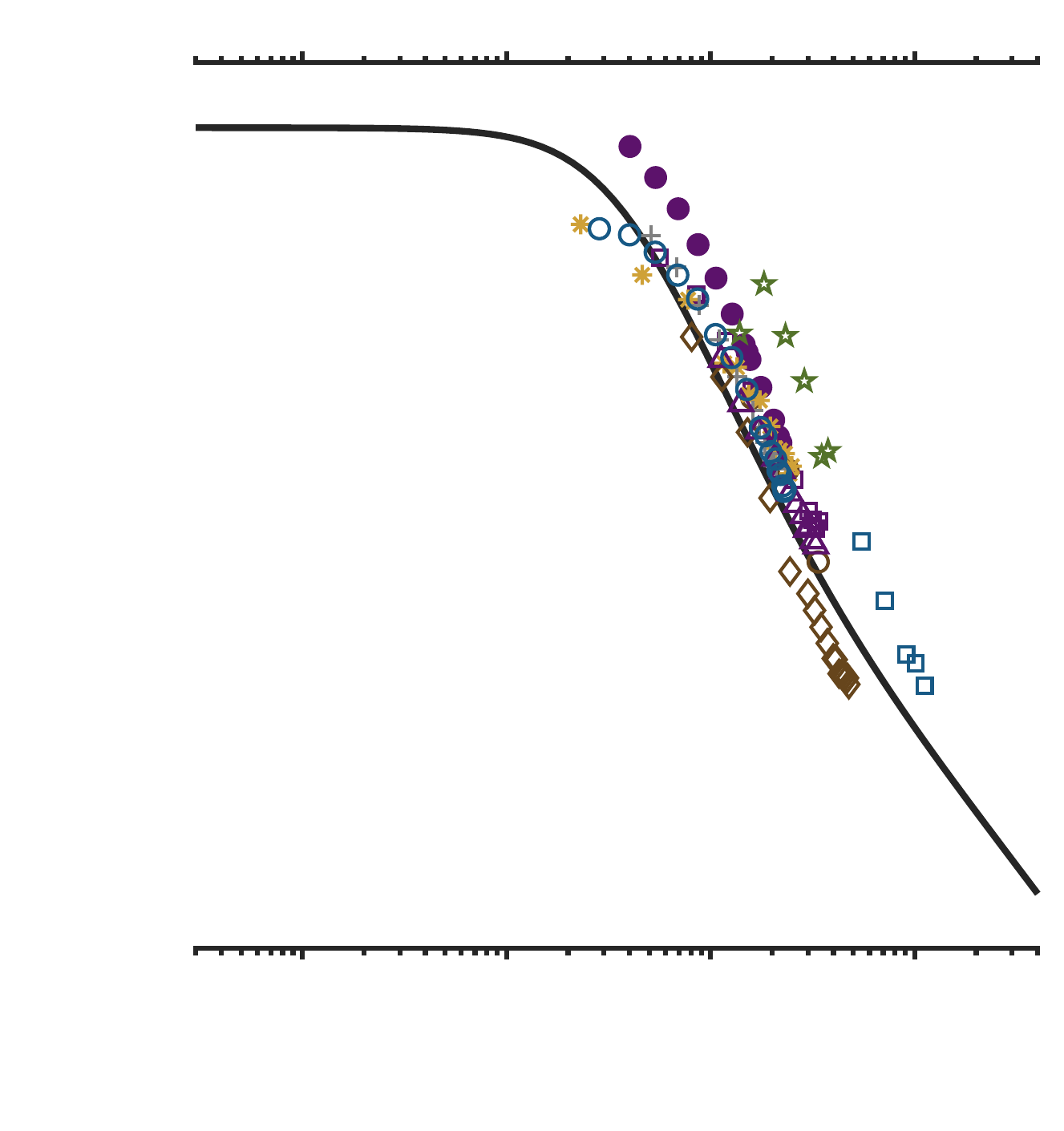}}%
    \put(0.25,0.1){\makebox(0,0)[lt]{\lineheight{1.25}\smash{\begin{tabular}[t]{l}$10^{-1}$\end{tabular}}}}%
    \put(0.45,0.1){\makebox(0,0)[lt]{\lineheight{1.25}\smash{\begin{tabular}[t]{l}$10^{0}$\end{tabular}}}}%
    \put(0.64,0.1){\makebox(0,0)[lt]{\lineheight{1.25}\smash{\begin{tabular}[t]{l}$10^{1}$\end{tabular}}}}%
    \put(0.84,0.1){\makebox(0,0)[lt]{\lineheight{1.25}\smash{\begin{tabular}[t]{l}$10^{2}$\end{tabular}}}}%
    \put(0.4,0.05){\makebox(0,0)[lt]{\lineheight{1.25}\smash{\begin{tabular}[t]{l}Cauchy Number $C_y$\end{tabular}}}}%
    \put(0,0){\includegraphics[width=\unitlength,page=2]{figure8pdf.pdf}}%
    \put(0.09,0.17){\makebox(0,0)[lt]{\lineheight{1.25}\smash{\begin{tabular}[t]{l}$10^{-1}$\end{tabular}}}}%
    \put(0.09,0.93){\makebox(0,0)[lt]{\lineheight{1.25}\smash{\begin{tabular}[t]{l}$10^{0}$\end{tabular}}}}%
    \put(0.09,0.4){\rotatebox{90}{\makebox(0,0)[lt]{\lineheight{1.25}\smash{\begin{tabular}[t]{l}Reconfiguration Number $R$\end{tabular}}}}}%
  \end{picture}%
\endgroup%

\caption {Contrasting the model proposed by \citet{gosselin2010drag} ($\textcolor[rgb]{0,0,0}{\rule{0.5cm}{0.5mm}}$) with experimental observations to assess the impact of flexibility on the drag of flexible plates.
Reconfiguration number $R$ obtained from the experimentally measured flexible and rigid drag force, is plotted against Cauchy number $C_y$. The markers represents different specimens: ($\textcolor[rgb]{0.40,0.27,0.11}{\circ}$) Yu140, ($\textcolor[rgb]{0.36,0.07,0.42}{\square}$) My140, ($\textcolor[rgb]{0.36,0.07,0.42}{\bullet}$) My100, ($\textcolor[rgb]{0.81,0.63,0.22}{\ast}$) Ye100, ($\textcolor[rgb]{0.50,0.50,0.50}{+}$) Gr120, ($\textcolor[rgb]{0.40,0.27,0.11}{\lozenge}$) Yu100, ($\textcolor[rgb]{0.33,0.45,0.17}{\star}$) Gre100, ($\textcolor[rgb]{0.09,0.35,0.52}{\square}$) Bl400, ($\textcolor[rgb]{0.36,0.07,0.42}{\triangle}$) My200, ($\textcolor[rgb]{0.09,0.35,0.52}{\circ}$) Bl80.}
\label{fig:RCY}
\end{figure}

These dimensionless numbers are plotted in Figure \ref{fig:RCY} for various specimens with different materials and dimensions. They are represented by the colour markers and calculated from the experimental data of the wind tunnel tests. The black curve represents the result found with the semi-empirical model of \citet{gosselin2010drag}.

From Figures \ref{fig:Fx vs Cvx Vs U Mylar} and \ref{fig:Fx vs Cvx Vs U Blue plastic}, we have shown that the drag force varies with the flow velocity. Moreover, the observed variation follows different laws when transitioning from a rigid to a flexible case, and it further depends on the mass number of the specimen. By dimensionless scaling of our force and velocity, this allows us to group all our specimens onto a single curve when there is static reconfiguration.
It is known that for a Cauchy number lower than 1, the reconfiguration number will be approximately equal to 1 and this defines the case of a rigid structure\cite{gosselin2010drag}.
The reconfiguration number emphasizes the effect of flexibility on the drag by comparing the flexible case to the rigid case, noticing that it decreases by increasing the Cauchy number. At large Cauchy number, $R$ approaches a power law $C_y^{-1/3}$ found by the dimensionless analysis in the static regime. Experimentally, we also found that this power is approximately equal to $-1/3$. Following these results, we can see that for a Cauchy number higher than 1, the reconfiguration becomes more important. In other words, the higher the flexibility, the larger the deformation.

At the onset of flutter, the reconfiguration number is expected to diverge from the static curve at a critical Cauchy number\cite{leclercq2018does}, $C_{y_{cr}}$. To illustrate these observations in more details, let us revisit the case of My140 presented above. Figure \ref{fig:RCY-max-My140} is similar to the previous one but focuses on a single case and ventures into the dynamic behaviour observed in the larger air velocity range. The values of the average reconfiguration number, $R$, and the maximum reconfiguration number, $R_{max}$, are presented in this figure. At a value around $C_{y_{cr}} \approx 3.4\times 10^1 $ indicated with a vertical dash line, flutter appears with a symmetrical mode of vibration. This is indicated with the (I) in Figure \ref{fig:RCY-max-My140}. The fluttering instability leads to a sudden increase of fluctuating drag $R_{max}$, while leaving the time-averaged drag unaffected to follow on the same static reconfiguration trend.
In the symmetrical flutter region, the maximum reconfiguration number $R_{max}$ reaches values significantly greater than 1, implying that the instantaneous drag fluctuations on the flexible specimen are markedly larger than the drag on the rigid specimen. At $C_{y_{cr}} \approx 6 \times 10^1 $ indicated with another vertical dash line the anti-symmetrical mode appears, we also observe a significant increase of both $R$ and $R_{max}$.

\begin{figure}
\centering
\normalfont
\def\svgwidth{\textwidth}
\input{figure9.tex}
\caption{Illustrating the onset of instability and its impact on the loss of drag reduction of the specimen My140. The plot displays both mean ($\textcolor[rgb]{0.6,0.22,0.22}{\circ}$) and maximum ($\textcolor[rgb]{0.6,0.22,0.22}{\rule{0.3cm}{0.5mm}}$) values (representing the $97^{th}$ percentile) of the reconfiguration number $R$ versus Cauchy number $C_y$. The black curve ($\textcolor[rgb]{0,0,0}{\rule{0.5cm}{0.5mm}}$) represents the prediction of $R$ by the model of \citet{gosselin2010drag}. Critical Cauchy number: $C_{y_{cr}} \approx 3.4\times 10^1$. (I) First vibration mode: symmetrical. (II) Second vibration mode: anti-symmetrical.}
\label{fig:RCY-max-My140}
\end{figure}

The drag measurements of another specimen, Bl400, are presented in Figure \ref{fig:RCY-max-Bl400} with the corresponding dimensional results in Figure \ref{fig:Fx vs Cvx Vs U Blue plastic}. Similarly to the previously considered specimen My140, the time-averaged reconfiguration number deviates only slightly from the static drag curve at the onset of instability at \(C_{y_{cr}} \approx 1.2 \times 10^2\) shown with the vertical dash line. It is however associated with a large increase in time fluctuations of the drag $R_{max}$. Note that for this specimen, the loss of stability directly leads to an anti-symmetric mode of vibration. The deviation of the static curve with the average number of reconfiguration is initially slight, then becomes more pronounced at \(C_{y} \approx 6.4 \times 10^2\). At this speed, during experiments, it was observed that both sides of the plate touched, resulting in a greater force in the \(x\)-direction. The cause of the deviation in the time-averaged drag load from the current tests done thus far remains unclear. It is uncertain whether this deviation is a consequence of the plates reaching their deformation limit or if it is associated with mechanical impacts occurring between the plates. Due to the high flexibility of Bl400 compared to My140, there is a significant deviation of the reconfiguration number from the static curve at higher Cauchy numbers compared to when flutter appears.

\begin{figure}
\centering
\normalfont
\def\svgwidth{\textwidth}
\input{figure10.tex}
\caption{Illustrating the onset of instability and its impact on the loss of drag reduction of the specimen Bl400. The plot displays both mean ($\textcolor[rgb]{0.6,0.22,0.22}{\circ}$) and maximum ($\textcolor[rgb]{0.6,0.22,0.22}{\rule{0.3cm}{0.5mm}}$) values (representing the $97^{th}$ percentile) of the reconfiguration number $R$ versus Cauchy number $C_y$. The black curve ($\textcolor[rgb]{0,0,0}{\rule{0.5cm}{0.5mm}}$) represents the prediction of $R$ by the model of \citet{gosselin2010drag}. Critical Cauchy number: $C_{y_{cr}}\approx 1.2 \times 10^2$. (II) Second vibration mode: anti-symmetrical.}
\label{fig:RCY-max-Bl400}
\end{figure}

These results highlight the impact of flexibility on drag reduction by static reconfiguration, on the variation of drag when flutter instability occurs and when there is a transition between the two vibration modes. We can map these boundaries using the following two dimensionless numbers: Reduced velocity $U_r$ and mass number $M^*$ . The reduced velocity corresponds to the free flow velocity made dimensionless by the stiffness and inertia of the rectangular plate: 
\begin{equation}
  U_r=UL\sqrt{\frac{m_s}{D}}.  
\end{equation}
The reduced velocity and the Cauchy number are related through the mass number as
\begin{equation}
    C_y=\frac{1}{16}C_D M^* U_r^2.
\end{equation}

In Figure \ref{fig:UrCyM}a, the variation of $U_r$ is presented for each specimen tested in the wind tunnel by changing $M^*$. The stability limit and mode boundaries are identified through three distinct approaches: firstly, using statistical analysis by computing the coefficient of variation (Figure \ref{fig:Fx vs Cvx Vs U Mylar} and \ref{fig:Fx vs Cvx Vs U Blue plastic}); secondly, employing a frequency domain analysis; and thirdly, relying on visual observations and data acquisition using the camera setup (Figure \ref{fig:pictures}). With these diverse datasets, we were able to determine the critical points for the onset of flutter and the transition between the two observed vibration modes. The mass number of this experimental campaign varies within the range of $0.1<M^*<6$. This interval has been set based on the materials used, while also avoiding significant 3D effects and edge torsion in the flexible plates. For all the experiments, a fixed value is defined for the aspect ratio, which is equal to $1/2$. By using a constant aspect ratio, it allows a more accurate and reliable comparison of results.

\begin{figure*}
\centering
\normalfont
\def\svgwidth{0.9\textwidth}
\input{figure11.tex}
\caption{a) Mapping of instability patterns and stability limit. The plot represents reduced velocity $U_r$ versus mass number $M^*$. b) Optimal flexibility from heavier to lighter rectangular plates. The plot represents Cauchy number $C_y$ versus mass number $M^*$. Green zone: reconfiguration. Blue zone: symmetrical vibration mode. Yellow zone: Anti-symmetrical vibration mode.}
\label{fig:UrCyM}
\end{figure*}

For low mass numbers, we observe flutter instability occurrence at high reduced velocity with a symmetric mode. During the experiments, we were unable to observe the anti-symmetric mode at high flow velocities for this cases. The first critical instability curve follows the least-square fit $U_{r_{I}}\approx 21 \times M^{*^{-0.4}}$ with a coefficient of determination $R=0.905$, which separates static reconfiguration from the symmetric mode. For mass numbers slightly superior to 1, we start to observe the anti-symmetric mode after mode $I$ is obtained. The transition between these two modes roughly follows the function $U_{r_{II}}\approx 26 \times M^{*^{-0.4}}$, $R=0.915$. For a mass numbers higher than $3\times10^0$, corresponding to the lightest cases, instability appears only with the anti-symmetric mode without passing through the symmetric mode. In summary, the findings indicate a dependence of the dynamic behaviour on the variation of mass number. Similar conclusion can be drawn from the stability maps presented by \citet{alben2008flapping, eloy2008aeroelastic,michelin2008vortex} for a flag. Their findings show  the appearance of lobes denoting different instability modes as $M^*$ increases and predicting the transitions between them. The first mode is observed for low mass numbers $M^* < 1$ and the transition to the second one occurs for values slightly above 1 and within the same range of reduced velocity values as in our case for a perpendicular plate.   

From the second mapping (Figure \ref{fig:UrCyM}b), we can identify the optimal flexibility that allows for maximum drag reduction, leading to improved aerodynamic performance without entering into instability, which has the opposite effect. It is noteworthy that a higher Cauchy number corresponds to greater deformation in our structure during the stable regime, leading to increased reconfiguration and, consequently, a reduction in drag. If we are seeking optimal flexibility based on the mass ratio, it can be found at the first boundary between reconfiguration and the first mode for $M^*< 3\times10^0$ and at the second boundary for $M^*>3\times10^0$.

\section{concluding remarks}
The static reconfiguration and flutter instability of  flexible structures in air was experimentally studied using rectangular plates clamped in their centre and positioned perpendicular to an airflow in the wind tunnel.
This study comprehensively examines the dynamic behaviour and drag variation of flexible plates under varying airflows. The analysis of the drag force variation between the rigid and flexible cases reveals that, while both experience an increase in drag with velocity, the flexible plate exhibits a less pronounced increase, resulting in drag reduction compared to the rigid case. However, beyond a critical velocity, the advantage of flexibility diminishes due to the onset of instability. In the post critical regime, two types of flutter were observed depending on the mass number $M^*$. For a low mass number, the instability occurs for high reduced velocities values and with a symmetrical vibration mode. For a mass number around 1, we observe a second vibration mode (anti-symmetrical). For lighter structures, we observe flutter instability at lower velocities, directly manifesting an anti-symmetrical mode. A detailed exploration of transitions from static to dynamic behaviour enables the mapping of stability limits and instability patterns. This mapping also aids in identifying optimal flexibility, allowing for drag reduction across a wide range of mass number values. The findings contribute valuable insights into the interplay of flexibility, instability, and aerodynamic performance in the dynamic behaviour of flexible plates. Nevertheless, further experimental investigations in the wake of the flexible plates are necessary to validate the presence or absence of VIVs. Another avenue to explore will involve varying boundary conditions to observe changes in vibration modes, shedding light on the impact of fixation.\\ 
Dynamic loss of stability imposes a limit to how much drag reduction is achievable through elastic reconfiguration in a slender structure. For real plant organs, such as leaves of branches or for whole plants, other limits might occur before a dynamic instability: pruning \cite{lopez2014drag} and material failure, contact and clumping \cite{vogel1989drag}.
Although we show here that flutter increases the dynamical load experienced by a thin sheet, rapid motion could also have beneficial effects in terms of thermoregulation \cite{vogel2009glimpses} and fending off herbivors.

\newpage
\section*{Acknowledgement}
The financial support of the Natural Sciences and Engineering Research Council of Canada  [funding reference number RGPIN-2019-7072] and the Canada Research Chair Program is acknowledged.

\section*{Appendix 1}
\label{sec:appendix1}

$\bullet$ Table \ref{tab:mast natural frequency} shows the caracteristics of the masts used in the experiments.

\begin{table}
\caption{\label{tab:mast natural frequency} Caracteristics of the masts used in the experiments.}
\begin{ruledtabular}
\begin{tabular}{cccc}
\textbf{Mast} & \textbf{Mast} & \textbf{Plate} & \textbf{Natural} \\
\textbf{Dia. ($mm$)} & \textbf{Length $L_m$ ($mm$)} & \textbf{Width $w$ ($cm$)} & \textbf{freq. ($Hz$)} \\
\hline
12.7 & 250 & 20-15 & 71-77 \\
12.7 & 180 & 14-9 & 128-148 \\
7 & 100 & 8-5 & 205-265 \\
7 & 65 & 4-1 & 584-626 
\end{tabular}
\end{ruledtabular}
\end{table}

$\bullet$ Table \ref{tab:specimen characteristic} presents the details of the tested specimens.

\begin{table*}
\caption{\label{tab:specimen characteristic} Rectangular specimen used in the wind tunnel tests, respecting an aspect ratio of 1:2.}
\begin{ruledtabular}
\begin{tabular}{cccccccc}
\textbf Specimen & {Plate} & \textbf{Plate} & \textbf{Plate} & 
  \textbf{Plate} & \textbf{Area Density} & \textbf{Mass} & \textbf{Flexural} \\
\textbf code & {Material} & \textbf{Thickness $t$ ($mm$)} &\textbf{Length $L$ ($mm$)} &\textbf{Width $w$ ($mm$)} & \textbf{ $m_s$ ($kg/m^2$)} & \textbf{Number $M^*$} & \textbf{Rigidity $D$ ($Nmm$)}  \\
\hline
My100 & Mylar & 0.178 & 100 &50& 0.2476 & 0.522 & 2.20 \\
My140 & Mylar & 0.178 & 140 &70& 0.2476 & 0.730 & 2.20 \\
My200 & Mylar & 0.178 & 200 &100& 0.2476 & 1.044 & 2.20 \\
My300 & Mylar & 0.178 & 300 &150& 0.2476 & 1.566 & 2.20 \\
My420 & Mylar & 0.178 & 420 &210& 0.2476 & 2.192 & 2.20 \\
Bl50 & Blue PET & 0.127 & 50 &25& 0.179& 0.360 & 1.10 \\
Bl80 & Blue PET & 0.127 & 80 &40& 0.179& 0.576 & 1.10 \\
Bl400 & Blue PET & 0.127 & 400 &200& 0.179& 2.880 & 1.10 \\
Gr120 & Gray PET & 0.190 & 120 &60& 0.241 & 0.644 & 3.40 \\
Ye50 & Yellow PET & 0.102 & 50 &25& 0.138 & 0.466 & 0.54 \\
Ye100 & Yellow PET & 0.102 & 100 &50& 0.138 & 0.934 & 0.54 \\
Re50 & Red PET  & 0.051 & 50 &25& 0.067 & 0.970 & 0.11 \\
Re100 & Red PET  & 0.051 & 100 &50& 0.067 & 1.940 & 0.11 \\
Re140 & Red PET  & 0.051 & 140 &70& 0.067 & 2.714 & 0.11 \\
Re200 & Red PET  & 0.051 & 200 &100& 0.067 & 3.876 & 0.11 \\
Re300 & Red PET  & 0.051 & 300 &150& 0.067 & 5.814 & 0.11 \\
Gre50 & Green PET  & 0.076 & 50 &25& 0.108 & 0.600 & 0.18 \\
Gre100 & Green PET  & 0.076 & 100 &50& 0.108 & 1.200 & 0.18 \\
Gre140 & Green PET  & 0.076 & 140 &70& 0.108 & 1.680 & 0.18 \\
Gre200 & Green PET  & 0.076 & 200 &100& 0.108 & 2.400 & 0.18 \\
Gre300 & Green PET  & 0.076 & 300 &150& 0.108 & 3.600 & 0.18 \\
Gre420 & Green PET  & 0.076 & 420 &210& 0.108 & 5.038 & 0.18 \\
Yu140 & Yupo  & 0.134 & 140 &70& 0.151 & 1.198 & 0.78 \\
Yu100 & Yupo  & 0.134 & 100 &50& 0.151 & 0.856 & 0.78 \\
\end{tabular}
\end{ruledtabular}
\end{table*}

$\bullet$ Figure \ref{fig:drag signal} presents the $97^{th}$ and $100^{th}$ percentiles of drag force over time. The choice of the 97th percentile in our study emphasizes a targeted approach, offering a robust perspective on data distribution, while the 100th percentile reflects the absolute maximum recorded drag force.

\begin{figure}
\centering
\normalfont
\def\svgwidth{\textwidth}
\input{figure12.tex}
\caption{ Variation of drag force ($F_x$) over time. Dashed lines represent the $97^{th}$ (\textcolor{red}{Red}) and $100^{th}$ (\textcolor{blue}{Blue}) percentiles of the drag force for the flexible rectangular plate YU140 at a specific velocity $U=10.7 \mathrm{m/s}$.}
\label{fig:drag signal}
\end{figure}

\section*{References}
\bibliography{aipsamp}
\bibliographystyle{unsrtnat}
\end{document}